\documentclass{aa}
\usepackage{graphicx}
\begin{document}
   \title{EVN observations of candidate Compact Symmetric Objects}


   \author{Liu Xiang
          \inst{1}
          \and
          C. Stanghellini\inst{2}        \and
          D. Dallacasa \inst{3,4}          \and
          Zhang Haiyan \inst{5} \
          }

   \offprints{C. Stanghellini}

   \institute{National Astronomical Observatories, Chinese Academy of Sciences, Urumqi 830011, China\\
           \email{liux@ms.xjb.ac.cn}
            \and
             Istituto di Radioastronomia del CNR, C.P. 141, I-96017 Noto SR, Italy\\
             \email{cstan@ira.cnr.it}
         \and
             Istituto di Radioastronomia del CNR, via P. Gobetti 101, I-40129 Bologna, Italy\
         \and
Dipartimento di Astronomia, Alma Mater Studiorum, via Ranzani 1, I-40127
Bologna,  Italy\\
             \email{ddallaca@ira.cnr.it}
         \and
National Astronomical Observatories, Chinese Academy of Sciences, Beijing 100012, China\\
\email{hyzhang@class1.bao.ac.cn}
             }
\date{Received September 14, 2001; accepted January 28, 2002}

   \abstract{We present pc-scale images of ten Compact Symmetric Objects (CSO) candidates
observed with the European VLBI network (EVN).
Five radio sources have been observed at 1.6 GHz, and five more at 2.3/8.4 GHz, the latter
subsample with the inclusion in the VLBI array of 3 antennae normally used for geodesy.
These objects were selected from existing samples of GHz Peaked Spectrum
(GPS) radio sources with the purpose to find and/or confirm the CSO
classification. These new VLBI observations allow us to
confirm the classification of two CSO candidates, and
to find a few new ones. The association of GPS radio galaxies with
a CSO morphology is strengthened by our findings, and this
result suggests an efficient way to increase  the
number of known CSOs by means of VLBI observations of
compact radio galaxies showing a convex radio spectrum.
\\
               }
   \maketitle

\section{Introduction}

Compact Symmetric Objects (CSOs) form a class of radio sources
with distinctive radio properties.
They are very powerful and compact sources with overall size $<$ 1 kpc, dominated
by lobe/jet emission on both sides of the central engine, and
are thought to be relatively free from beaming effects (Wilkinson et al. 1994). Their small
size is likely due to their youth ($<$ 10$^4$ yr).
This hypothesis has been largerly accepted after the detection and the estimate
of the separation speed of the micro hot-spots in a couple of CSOs
(Owsianik and Conway 1998, and Owsianik et al. 1998).
Since then, expansion velocities of the outer edges
have been detected or suspected in a handful of other CSOs (Fanti 2000 and
references therein).

A unification
scenario assumes that CSOs evolve into Medium-size
Symmetric Objects (MSO), which, in turn, evolve into Large Symmetric
Objects (LSO), i.e. large FRII radio sources (Fanti et al. 1995, Readhead et al. 1996ab,
Snellen et al. 2000).
The monitoring of the structural changes over time seems one
of the more promising ways to understand the evolution of these objects,
and a larger number of CSOs suitable for these repeated observations is desirable.

Most confirmed CSOs and
CSO candidates are commonly found to have a global convex radio spectrum
peaking at GHz frequencies, thus belonging to the class of the GHz Peaked Spectrum (GPS)
radio sources.

These latter objects are known to show structures resolved only with VLBI. Only about 10\% of
them shows faint extended radio emission on scales
of a few tens of kpc, while $\sim$90\% of them is entirely
contained within the extent of the narrow-line region ($<$ 1 kpc) of the host
galaxy (Stanghellini et al. 1998).
GPS radio sources are usually identified with galaxies at low or intermediate redshifts (z$<$ 1), and
quasars at higher redshifts (O'Dea et al. 1991, Snellen et al. 1999), and they make up a
significant fraction  ($\sim$10\%) of the bright centimeter wavelength
selected radio sources.

Many GPS radio sources have
no optical counterpart yet, and these empty fields are most
likely distant galaxies, too faint to be detected. Stanghellini et al. (1997a, 1998) find a strong
correlation between the milliarcsecond (mas) morphology and the optical host
in bright GPS sources. Galaxies are generally associated
with CSOs while quasars have more often core-jet or complex morphology.

There are a few tens of sources in known GPS samples, and many more are
in candidate lists; several of these sources have either no or poor VLBI images.
In order to increase the number of
known CSOs, we selected from the lists of known bright GPS samples (O'Dea et al. 1991,
de Vries et al. 1997, Stanghellini et al. 1998, Dallacasa et al. 2000)
the sources accessible to the EVN without enough structural information
in the literature to permit a proper classification yet.
Scheduling the radio sources to be
observed in the allotted time we gave a higher priority to objects
with known double morphology at a single frequency.

We report here
the results of the two first observing runs.
In a first session at 1.6 GHz
with the EVN and MERLIN we observed 5 sources. At this frequency we could detect the presence
of extended steep spectrum emission, determine the total angular size
of the source, and possibly determine the low frequency spectral shape
of the components, when other images at higher frequencies were available.
Five further  radio sources were observed
at 2.3/8.4 GHz with the EVN plus 3 antennae commonly used in geodesy experiments.
These simultaneous observations at 2 frequencies are best suited to detect the cores.

\section{Observations and data reduction}

\subsection{EVN + MERLIN at 1.66 GHz}
EVN Observations were carried out in two separate VLBI experiments.
The first observation has been done on 13 November 1999 at 1.66 GHz
using the MKIII recording system, with a total bandwith of 28 MHz, in left circular
polarization, with the antennae of Effelsberg, Westerbork, Jodrell, Cambridge,
Onsala, Medicina, Noto, Torun, Nanshan and Sheshan. Snapshot
observations for a total of 5.2 hours of observing time were made for
the sources 1345+125, 1404+286 (OQ208), 1518+047, 1604+315, and 1751+278.
The MERLIN array simultaneously observed those sources with 7 antennae
(Defford, Cambridge, Knockin, Darnhall, MK2, Lovell and Tabley) at the same frequency.
The EVN 1.66 GHz data were correlated at the MPIfR MKIII correlator.

The MERLIN data were calibrated in a standard way
and the images have been made  using the DIFMAP  package (Shepherd 1997).
We only show the MERLIN image of 1345+125 as the other objects are
pointlike at the resolution of about 150 mas.
The MERLIN images of  the remaining sources are shown in Zhang et al. (2001).

\subsection{EVN + GEO at 2.3 and 8.4 GHz}
The second observation has been done on 29 May 2000 at 2.3/8.4 GHz
using the MKIV recording system with a bandwith of 16 MHz in each frequency, in right circular
polarization. The EVN antennae involved in this experiment which gave
useful data were Effelsberg, Westerbork,
Onsala, Medicina, Noto, Yebes, plus the geodetic antennae of
Matera, Ny-Alesund, Wettzel.
Nanshan and Sheshan failed to give fringes because of problems in their
recent upgrade to MKIV formatting system.
Snapshot
observations for a total of 7.9 hours of observing time were taken for
1333+589, 1427+109, 1509+054, 1526+670, and 1734+508.
The 2.3/8.4 GHz data were correlated at the MKIV correlator at JIVE.
This is one of a very few EVN experiment making use of several
geodetic antennae, and their inclusion in the array has proven
to be very useful with a major improvement in the NS
resolution and UV coverage.\\


The Astronomical Image Processing
System (AIPS) developed by the National Radio Astronomy Observatory (NRAO), has
been used for editing, a-priori calibration, fringe-fitting, imaging and self-calibration.
Several iterations of phase self-calibration were performed until we
converged to an acceptable solution. Then one or more iterations of amplitude and
 phase-calibration were applied. We also used the IBLED and CLIP tasks in AIPS for
editing the data before the first phase-calibration and/or after the first amplitude
calibration because of the dispersion of the data on some baselines.
The main source of errors of the flux densities estimated in our final
VLBI images is
the absolute flux density scale which is assumed to have uncertainties of $\sim$10\%.

The sizes and the flux densities of the components given in Tables 2 and 3
have been obtained with a gaussian fit using the task JMFIT in AIPS.

\begin{table*}
\caption{ GPS sources. Columns 1 through 13 provide: name, optical identification, optical
magnitude,  redshift (those with * are photometric estimates by Heckman et al. 1994), linear scale factor pc/mas
[H$\rm _o$=100 km s$^{-1}$ and q$\rm _o$= 0.5 have been assumed], maximum VLBI angular size, maximum
VLBI linear size,  flux density at 1.6 GHz as measured by the MERLIN, flux density
at 1.6, 2.3, 8.4 GHz as measured from our VLBI images, low frequency spectral index, high frequency
spectral index, computed turnover frequency, and references for the spectrum information
(we use S $\propto$ $\nu$$^{-\alpha}$):
[1] Stanghellini et al. 1998, [2] de Vries et al. 1997, [3] Dallacasa et al. 2000.}
\begin{flushleft}
\begin{tabular}{llllccrcccccc}
\hline
\hline
\noalign{\smallskip}
Source & id & m & z & scale & $\rm  \theta_{max}$ & $\rm l_{max}$
& $\rm S_{merlin}$ & $\rm S_{1.66}$ &  $\rm \alpha_{l}$ & $\rm \alpha_{h}$ & $\rm \nu_{m}$ &ref.\\
 & & & & $\rm pc/mas$ & mas & pc & Jy &Jy & & & GHz\\
\noalign{\smallskip}
\hline
\noalign{\smallskip}
 1345+125& G & 15.5r & 0.122 &1.5 & 100  & 150  & 4.88 & 4.33  & -0.67 & 0.50 & 0.4   & 1  \\
 1404+286& G & 14.6r & 0.077 &1.0 & 10   & 10   & 1.14 & 1.00  & -1.5  & 1.6  & 4     & 1  \\
 1518+047& Q & 22.2R & 1.296 &4.3 & 160  & 688  & 3.55 & 3.00  & -0.20 & 0.93 & 0.9   & 2  \\
 1604+315& G & 22.7r & 1.5*  &4.3 &$<$5  &$<$22 & 0.83 & 0.75  & -0.59 & 0.28 & 1.5   & 2  \\
 1751+278& G & 21.7R & 0.86* &4.2 & 50   & 210  & 0.61 & 0.53  & -0.27 & 0.57 & 1.4   & 2  \\
\noalign{\smallskip}
\hline
\noalign{\smallskip}
  &  &  &  &  & & & $S_{2.3}$ & $S_{8.4}$ & &  &  & \\
 \noalign{\smallskip}
\hline
\noalign{\smallskip}
1333+589&EF&       &       &    & 15   &      &  0.57    & 0.58 & -0.84 & 0.52 & 4.9   & 3 \\
1427+109&Q & 18.5B & 1.71  &4.2 & $<$2 &$<$8.4&  0.70    & 0.89 & -1.18 & 0.55 & 4.9   & 3 \\
1509+054&G & 16.2  &       &    & 10   &      &  0.20    & 0.71 & -1.72 & 0.46 & 11    & 3 \\
1526+670&Q & 17.2E & 3.02  &3.6 & 5    & 18   &  0.25    & 0.28 & -1.26 & 1.21 & 5.8   & 3 \\
1734+508&G?& 23.1R &       &    & 8    &      &  0.71    & 0.81 & -0.5  &  0.4 & 5.9   & 3 \\
\noalign{\smallskip}
\hline
\end{tabular}
\end{flushleft}
\end{table*}

\begin{table*}
\caption{ Parameters of the components in the VLBI images at 1.66 GHz. The columns
give: (1) source name and possible classification (CSO: Compact Symmetric
Object; cj: core-jet), (2) component identification, (3) and (4) major and minor axes when a fit
has been possible
(using JMFIT in AIPS), (5) position angle of the component, (6) and (7) distance and
position angle of component with respect to the first component, (8) total flux density of the
component, (9) intrinsic brightness temperature neglecting relativistic effects, (10) minimum
energy density, (11) equipartition magnetic field.
\hfill\break
}
\begin{flushleft}
\begin{tabular}{clcccccccccccc}
\hline
\hline
\noalign{\smallskip}
Source and & & $\theta_1 $ & $\theta_2 $ & PA & R & PA &
$\rm S_{1.66}$ & T$\rm _b$ & u$\rm _{min}$ & H$\rm _{eq}$     \\
class       &       & mas       & mas     & $^\circ$    &  mas  & $^\circ$ &   mJy    &
$\rm 10^9K^\circ$   &$\rm 10^{-6}erg/cm^3$& $\rm 10^{-3}G$        \\
\noalign{\smallskip}
\hline
\noalign{\smallskip}
 1345+125 & A     &      &      &     &      &       &   $<$40   &         &      &         \\
     CSO  & B     & 8    & 1.7  &136  &  0   &       &     302   &  5    & 14   &   12    \\
          & C     & 13   & 4.5  &169  & 8.2  &  138  &     383   &  1.4    & 4  &   7       \\
          & D     &      &      &     &      &       &    3592   &         &       &        \\
          & E     &      &      &     &      &       &      45   &         &       &        \\
1404+286  & NE    & 1.9  & 1.2  & 93  &  0   &       &     958   &  90     & 100   &  30    \\
    CSO   & SW    & 2.7  & 1.3  & 2   & 7.8  & -115  &    14     & 0.8     & 7   &  9   \\
1518+047  & A     & 11   & 3.4  & 33  & 0    &       &    1675   &  20     &  35   &  20    \\
  CSO?    & C     & 5.7  & 4.5  & 50  &136.8 & -153  &    1167   & 20      &  30   &  20    \\
1604+315  &       &$<$5  &$<$5  &     &      &       &     746   & $>15$   &$>25$  & $>20$  \\
1751+278  & A     & 2.8  & 1    &92   & 0    &       &     479   &  70     & 110   &  35    \\
  cj      & B     & 3.7  & 3.7  & 1   & 22   & -122  &      39   &  1.0    & 4     &  7   \\
          & C     & 17   & 3.2  &101  & 23   & -120  &      13   &  0.09   &  1.1  &  4   \\
          & D     & 17   & 4.6  &152  & 43   & -94   &       3   &  0.01   &  0.3  &  2   \\
\noalign{\smallskip}
\hline
\noalign{\smallskip}
\end{tabular}
\end{flushleft}
\end{table*}

\begin{table*}
\caption{ Parameters of the components in the VLBI images at 2.3 and 8.4 GHz. The columns give: (1)
source name and possible classification (CSO: Compact Symmetric Object, cj: core-jet; d: double), (2)
component identification, (3) total flux density of the
component at 2.3 GHz, (4) and (5) major and minor axes at 2.3 GHz when fit has
been possible (using JMFIT in AIPS), (6)
position angle of the component at 2.3 GHz, (7) total flux density of the component at 8.4 GHz, (8)
and (9) major and minor axes at 8.4 GHz, (10) position angle of the component at 8.4 GHz,
(11) and (12) distance and position angle of component with
respect to the first component at 8.4 GHz, (13), (14)
and (15) intrinsic brightness temperature neglecting
relativistic effects, minimum energy density, equipartition
magnetic field at 8.4 GHz.
\hfill\break
}
\begin{flushleft}
\begin{tabular}{clcccccccccccccccc}
\hline
\hline
\noalign{\smallskip}
Source and &  &S$_{2.3}$  & $\theta_1 $ & $\theta_2 $ & PA &
S$_{8.4}$ & $\theta_1 $ & $\theta_2 $ &PA & R & PA &  T$\rm _b$ &u$\rm _{min}$ & H$\rm _{eq}$     \\
class       &   &   Jy    & mas       & mas     & $^{\circ}$
 &Jy  &  mas   & mas  & $^{\circ}$ & mas   & $^{\circ}$&$\rm {10^9K^\circ}$&${10^{-6}\rm
 {erg\over cm^3}}$& ${10^{-3}\rm G}$ \\
\noalign{\smallskip}
\hline
\noalign{\smallskip}
 1333+589 & N  &0.16 &  4.0 & 1.7  & 153  &0.42 &  0.5 & $<$0.5 & 177 & 0     &      &      &    &    \\
  d/CSO?    & S  &0.41 &  2.1 & 1.2  & 176  &0.15 &  1.1  & 0.7   & 39  & 12.8  & -164 &      &    &    \\
 1427+109  &    &0.70 & $<$5 &$<$5  &      &0.89 & $<$2  & $<$2  &     &       &      &$>5$  &$>330$&$>60$    \\
 1509+054  & E  &0.20 &  4.8 & 1.2  &109   &0.3  &  0.8  &$<$0.5 & 98  &  0    &      &      &     &    \\
   d/CSO?   & W  &     &      &      &      &0.4  &  0.9  & 0.5   & 94  & 4.9   & -90  &      &     &    \\
 1526+670  & E  &0.25 &  1.9 & 1.3  & 82   &0.1  &  2    & 1     & 45  &  0    &      &  1.5 & 530  & 80\\
   cj?      & W  &     &      &      &      &0.17 &  0.8  & 0.8   &  1  & 0.3   & -126 &  8   &1600 & 130 \\
 1734+508   & N  &0.70 &  3.1 & 2.4  &   1  &0.15 &  0.8  & 0.5   & 157 &  0    &      &      &    &    \\
   d/CSO?   & S  &     &      &      &      &0.66 &  0.9  & 0.5   & 13  & 2.6   & -144 &      &    &   \\
\noalign{\smallskip}
\hline
\noalign{\smallskip}
\end{tabular}
\end{flushleft}
\end{table*}

\section{Results and comments on individual sources}

In this section we present the results from our VLBI observations. Basic information
on the target sources is presented in Table 1. Table 2 and Table 3 summarize the
parameters derived from the images. In the following we present each source and comment on
the results of the present observations.

\subsection{Images at 1.66 GHz}

\subsubsection{1345+125 (J1347+1217, 4C12.50)}

1345+125 at z=0.122, is one of the nearest bright GPS radio sources.
The radio source is associated with a galaxy (m$\rm _r$=15.5)
lying at the center of a group. The optical image shows 2 well-resolved nuclei
separated by 2" (4 kpc), with the radio source associated
with the westernmost one (Stanghellini et al. 1993, Evans et al. 1999).
The double nucleus, the irregular isophotes, the presence of a tail, and a number of close smaller companions
indicates that 1345+125 is likely to be in process of a merging, which had triggered the
radio activity (Heckman et al. 1986).

The 5 GHz radio image (Stanghellini et al. 1997a) shows a well defined structure visible
in Fig. 1a, where component A is
tentatively identified as the core, B and C are prominent knots in a jet,
D is a mini-lobe, and E is tentatively identified as a much weaker counter lobe
or jet. The whole radio emitting region is confined in less
than 150 pc. Our image is the first 1.66 GHz VLBI image
of 1345+125. We can identify the jet components (B, C), and the lobes (D, E), while
the candidate core (A) is not present, likely self absorbed. For comparison, we convolved the 5 GHz image
in Stanghellini et al. (1997a) and the 2.3 GHz image in Fey et al. (1996) with the restoring beam
of our 1.66 GHz image, and show
the three images in Fig.1a, superimposed as contours to the same full resolution
5 GHz grey scale image.
The image registration has been performed on the basis of the position of
the southernmost component, likely free from optical depth effects and with the hypothesis
that its proper motion relative to the other components B and C is small (like in the hot spots of CSOs).

We find that A becomes
weaker at 2.3 GHz and vanishes at 1.66 GHz. We
set a very conservative upper limit of 40 mJy for the flux density of this component at 1.66 GHz.
Component A has therefore an inverted spectrum, is most probably the core and hence 1345+125 is a CSO.

The source in our MERLIN image seems slightly resolved in the N-S direction (Fig. 1b).
A fit with a gaussian on the image gives a size of about 50 mas. Considering that the
angular size measured on an extended region is roughly twice the half maximum width
given by the gaussian fit, it is consistent with the size seen in the EVN image.\\

\subsubsection{1404+286 (J1407+2827, OQ208, Mkn668)}

The compact radio source at a redshift of 0.077 is one of the closest bright GPS galaxies.
With m$\rm _r$=14.6, the optical image presented by Stanghellini et al. (1993) shows the presence of
companions in the galactic envelope and a tail of low brightness emission in N-S direction, suggesting
that the galaxy is dynamically disturbed.

The radio source has been considered in terms of a CSO structure
(Stanghellini et al. 2000). It has a
weak core at 15 GHz, two-sided faint and short jets and double asymmetric
mini-lobes located within 10 mas, giving a
projected size of less than
10 pc. The two micro hot-spots of this radio source seem to increase their distance at
a rate of around 0.15{\it c}, comparable to what already found in other CSOs, yelding a dynamical age
of a few centuries (Stanghellini et al. 2000).
This radio source has been extensively studied at VLBI resolution,
but we included it in our observations at 1.66 GHz because of its remarkable
spectral properties at lower frequencies.
Actually, Kameno et al. (2000) report a very inverted spectrum below
the peak frequency of the weaker south-western region
of this source based on VSOP observations. They consider the extremely inverted
spectrum inconsistent with synchrotron self-absorption (SSA), and propose a process
of free-free absorption (FFA) to explain the radio spectral shape.
Our 1.66 GHz image (Fig. 2a) shows two components, a strong NE one (958 mJy)
and a weak SW one (14 mJy). Although there is a dirty beam sidelobe close to
the position of the southwest component, adding a source
of uncertainty to its flux density measure, our estimate of the lobe flux densities
from ground VLBI observations is consistent with that of Kameno et al. (2000).

We show the spectra of the two regions resolved by our observations in
Fig. 2b where the values at 2.3/8.4 and 5 GHz are from Fey et al. (1996) and
Stanghellini et al. (1997b) respectively.
We get a slope of the SW component between 1.6 and 2.3 GHz of
$\alpha$=-3.8$\pm$0.7, confirming a spectral shape for this component incompatible with simple SSA.

Our MERLIN image gives a pointlike source with no evidence of
diffuse emission at the MERLIN scale at a level of 3 $\times$ the r.m.s. noise of 3 mJy.\\

\subsubsection{1518+047 (J1521+0430, 4C04.51)}

This object at redshift 1.296 is identified with a quasar of magnitude
m$\rm _{R}$=22.2 (Stickel \& K\"uhr, 1996). 1518+047 is a strong radio source with
P$\rm _{5GHz}$=10$^{27.9}$ W/Hz, and is one of the first examples of a double source at mas resolution.
Its morphology is discussed
by Mutel et al. (1985), Phillips \& Mutel (1982). They imaged the source with the U.S. VLBI Network
at 1.67 GHz, 5 GHz and 610 MHz (Mutel et al. 1985, Mutel \& Hodges, 1986). At 5 GHz, the
source is resolved into two micro lobes, and each lobe further resolved
into two/three components. Dallacasa et al. (1998) also
resolve the source in two lobes at 2.3/8.4 GHz EVN observations. Our 1.66 GHz EVN image (Fig. 3a)
also shows two lobes, with a hint of a jetlike emission (D) associated with
the component C, also visible at other frequencies. The resolution of our data did not allow
to model components B and D convincingly, but we fitted the main components A and C as reported in
Table 2.
The spectra of the 2 components are pretty similar (Fig. 3b) and this
source is very likely a CSO even if a compact core has not been detected yet.
The MERLIN image which has a rms noise of 8 mJy/beam does not reveal any additional emission.\\

\subsubsection{1604+315 (J1606+3124)}

A galaxy with m$\rm _r$=22.7 at redshift 1.5 estimated photometrically
(as reported by Heckman et al. 1994), is close to
the radio position; although very faint, it shows a complex optical morphology elongated
in the E-W direction. Other faint objects are present a few arcseconds away, suggesting
1604+315 is a member of a cluster (Stanghellini et al. 1993). Our VLBI image shows no details (Fig.4).
This source is present in the VLBA calibrator list (Peck et al. 1998) showing a possible core-jet morphology.
No extended
emission has been detected in our MERLIN image with a rms noise of 1.6 mJy/beam.\\

\subsubsection{1751+278 (J1753+2750)}

This object is identified as a galaxy with m$\rm _R$=21.7 and redshift 0.86 estimated
photometrically (as reported by Heckman
et al. 1994), it is faint and extended (O'Dea et al. 1990), with
a possible companion (O'Dea et al. 1996).

No previous VLBI morphology
information is available for this source. Our VLBI image at 1.66 GHz (Fig. 5) exhibits a compact
dominant component (A), a secondary compact component to south-west (B), from which
a trail of radio emission points to west (C), then probably to north-west, where just a hint
of emission is visible (D).
No extended emission
has been detected in our MERLIN image with rms noise of 1.8 mJy/beam.

This source is classified as a GPS source on the basis of a flux density limit of 0.25 Jy
at 408 MHz (Spoelstra et al. 1985). In the literature three values
of the flux density of this source at 1.4 GHz are found,
which are 0.52$\pm$0.06 Jy (Condon et al. 1983, and references therein),
0.647$\pm$0.030 Jy (White et al. 1992), and 0.6253$\pm$0.0188 Jy (NVSS value). The flux density
at 365 MHz is 0.685$\pm$0.069 Jy (Douglas et al. 1996), so the flux density limit
at 408 MHz seems wrong, (or the source is highly variable at low frequency).
Given the spectral flattening of this source above 1.6 GHz, the lack of a
peak in the spectrum, and its possible
variability, the source could be
a flat spectrum or a compact steep spectrum (CSS), rather than  a GPS radio source
and we will not consider it in the discussion of the results.
\\

\subsection{Images at 2.3 and 8.4 GHz}
\subsubsection{1333+589 (J1335+5844, 4C58.26)}

This radio source has a peak
flux density of 0.75 Jy at 4.9 GHz; In the optical it
is an empty field (Stickel \& K\"uhr, 1994).
This object belongs to the CJ1 sample (the first Caltech--Jodrell
Bank survey), and has been already imaged with VLBI at 1.67 GHz (Thakkar et al. 1995) and 5 GHz
(Xu et al. 1995). The source is clearly resolved into two compact components at both forementioned
frequencies, and this morphology is confirmed in our 2.3/8.4 GHz images.
As shown in Fig. 6a and Fig. 6b, the two components are separated by 12.8$\pm$0.1 mas, and no jetlike feature
has been detected between them. We show in Fig. 6c the spectra of the components
using our images and the data at 1.67 and 5 GHz mentioned above.

The northern component
show a rising spectrum at all frequencies, indicating it is more compact. It
is possible that it is the core of the radio emission and this object being
a core-jet radio source. Indeed this source is not considered a CSO in Taylor (1996),
Polatidis et al. (1999), and Peck \& Taylor (2000),
who have examined all candidates in the PR, CJ1 and CJ2 surveys with
multi-frequency VLBA observations.
We think it is also possible that the rising
spectrum reflects a very compact micro-hotspot, smaller than in
the southern lobe, and the radio source still would be a CSO.

More sensitive high resolution observations to detect any further emission should provide a
better constraint for the correct morphological classification.

\subsubsection{1427+109 (J1430+1043)}

The optical host is a quasar at z=1.710 and magnitude m$\rm _B$=18.5 (NASA/IPAC Extragalactic Database).
In the radio band this source peaks at 4.9 GHz with
a flux density of 0.91 Jy. The source is not resolved in our 2.3/8.4 GHz images (not shown), and no previous
VLBI morphology information is available in the literature.

\subsubsection{1509+054(J1511+0518)}

The optical identification for this radio source is a galaxy with visual magnitude 16.2
(Dallacasa et al. 2000). The source shows
a very steep spectrum with a peak flux density of 0.77 Jy at 11 GHz.
Our 8.4 GHz image (Fig. 7) shows two components separated by 4.9 mas;
the resolution at 2.3 GHz does not allow to separate these two components.
No VLBI image is available in the literature for this source.
The double morphology at 8.4 GHz makes this source a CSO candidate.

\subsubsection{1526+670 (J1526+6650)}

The host of the source is a quasar at high redshift (z=3.02) and magnitude m$\rm _r$=17.2
(Hook et al. 1996). This is a CJ2 source and
has been imaged at 5 GHz (Taylor et al. 1994). It is slightly elongated in
our 8.4 GHz image (Fig. 8) in agreement with the 5 GHz image, and the radio emission
can be modelled with two components
as listed in Table 3. This object is unresolved in our 2.3 GHz image.

The spectrum of the components between 8.4 GHz and 5 GHz
(Taylor et al. 1994), shows that the western, brighter component has a
flatter spectral index than the
eastern component, hence it is possible that this source
is a core-jet object, but new observations at higher frequency
and resolution are necessary to give a proper classification.

\subsubsection{1734+508 (J1735+5049)}

A very faint (m$\rm _R$=23.1) optical object has been detected at the radio position, and tentatively
classified as a galaxy (Stickel \& K\"uhr, 1996). The radio source peaks at 5.9 GHz with a peak
flux density of 0.99 Jy. This is a CJ1 source, resolved in two components at 5 GHz, and
modelled with three gaussian components (Xu et al. 1995), Our 8.4 GHz image (Fig. 9)
shows a double morphology, while it is unresolved in our image at 2.3,
as at 1.6 GHz (Thakkar et al. 1995). Considering the
model given by Xu et al. at 5 GHz, the northern component has a steep spectrum ($\alpha \sim 0.8$),
while the southern one is slightly inverted ($\alpha \sim -0.2$). As for 1333+589 this source has
not been considered a CSO by Taylor (1996),
Polatidis et al. (1999), or Peck \& Taylor (2000). Also in this case we think it is possible
that an inverted
spectrum is consistent either with a core or a very compact hot-spot. Until additional
high resolution morphological information will be available we consider this source
a CSO candidate, given its double morphology.

\section{Discussion and conclusions}

We can summarize the results from
our EVN observations at 1.66, 2.3 and 8.4 GHz for 10 GPS radio sources as follows:

i) we confirm the CSO nature of
1345+125 revealing a component which is very likely to be the core.

ii) for the radio source 1404+286 we confirm from ground based VLBI observations
the findings by Kameno et al. (2000) i.e. that the radio spectrum is affected
by FFA rather than SSA, or maybe both. This is an important observational clue because
it implies the presence of a dense ionized ambient medium somewhere
between us and the strongly absorbed radio emitting region.

iii) we also found more CSO candidates which need need further
observations to be confirmed.

The relationship between radio mas morphology and optical host is rather
respected in the small sample considered here, although also quasars may appear in the CSO
candidate list (e.g. 1518+047).
Excluding 1751+278 which does not show a convex radio spectrum and
it has been erroneously classified as a GPS radio source, we find
that 5 galaxies (we include J1335+5844 in the galaxy group as empty fields
are mostly identified with
weak galaxies once sensitive optical observations become available)
are confirmed or candidate CSO, one quasar is also a CSO candidate
and 2 quasars are unresolved or slightly resolved objects in our images.
Therefore the observations of  GPS radio galaxies
seem to be an efficient way to find CSOs.

The MERLIN data at 1.66 GHz do not reveal any extended emission in four of the five sources
observed. Only 1345+125 is very sligthly resolved, in agreement with the size seen
in the EVN image.

Since the cores are known to have highly inverted spectra, multi-frequency
and especially high frequency sensitive
observations are needed to confirm the CSO nature
for the new CSO candidates discovered here. If they are
confirmed they will be very suitable objects to monitor
for separation speed, due to their small size (thus possibly
faster separation velocity) and the presence of strong and compact
components (hot-spots?) at their edges.

\begin{acknowledgements}
Part of this work was supported by the National Natural Science Foundation of China. L.X. wish
to thank finacial supports from the CNR Istituto di Radioastronomia (Italy) and the Joint
Institute of VLBI in Europe (in The Netherlands) for the work.  
\end{acknowledgements}

\clearpage
 \begin{figure*}
   \centering
   \includegraphics[width=12cm]{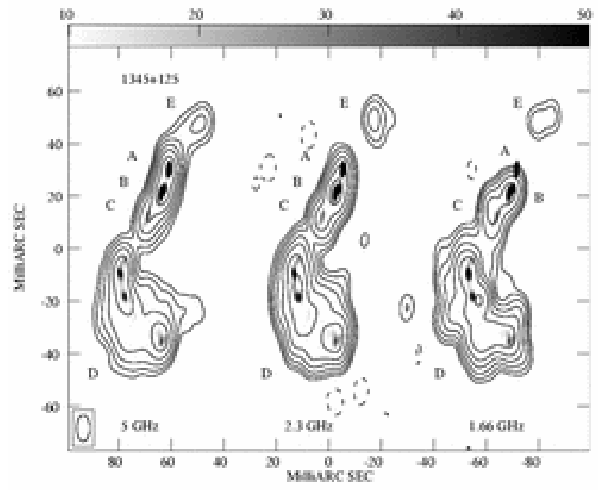}
     \includegraphics[width=8cm]{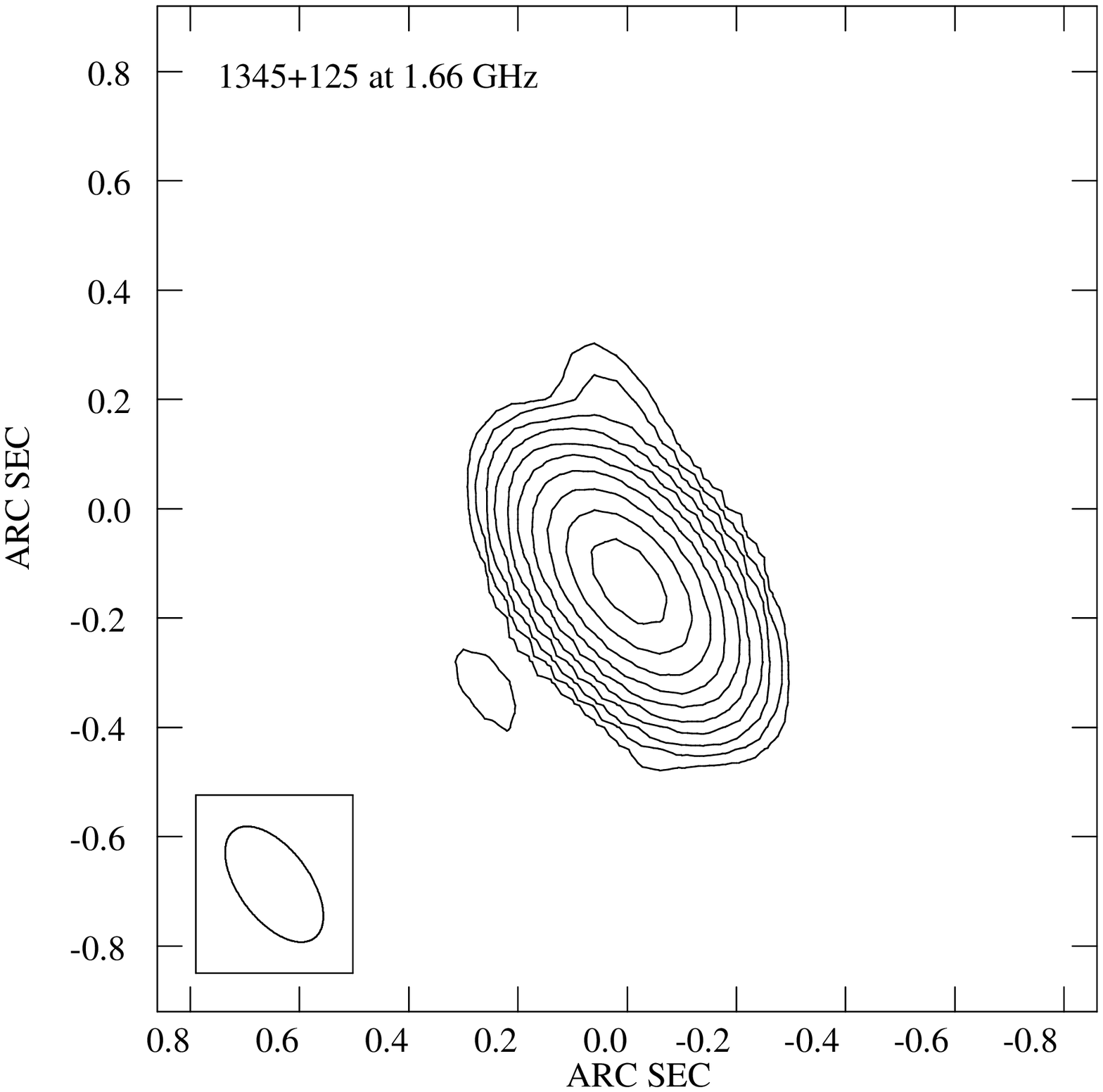}
      \includegraphics[width=8cm]{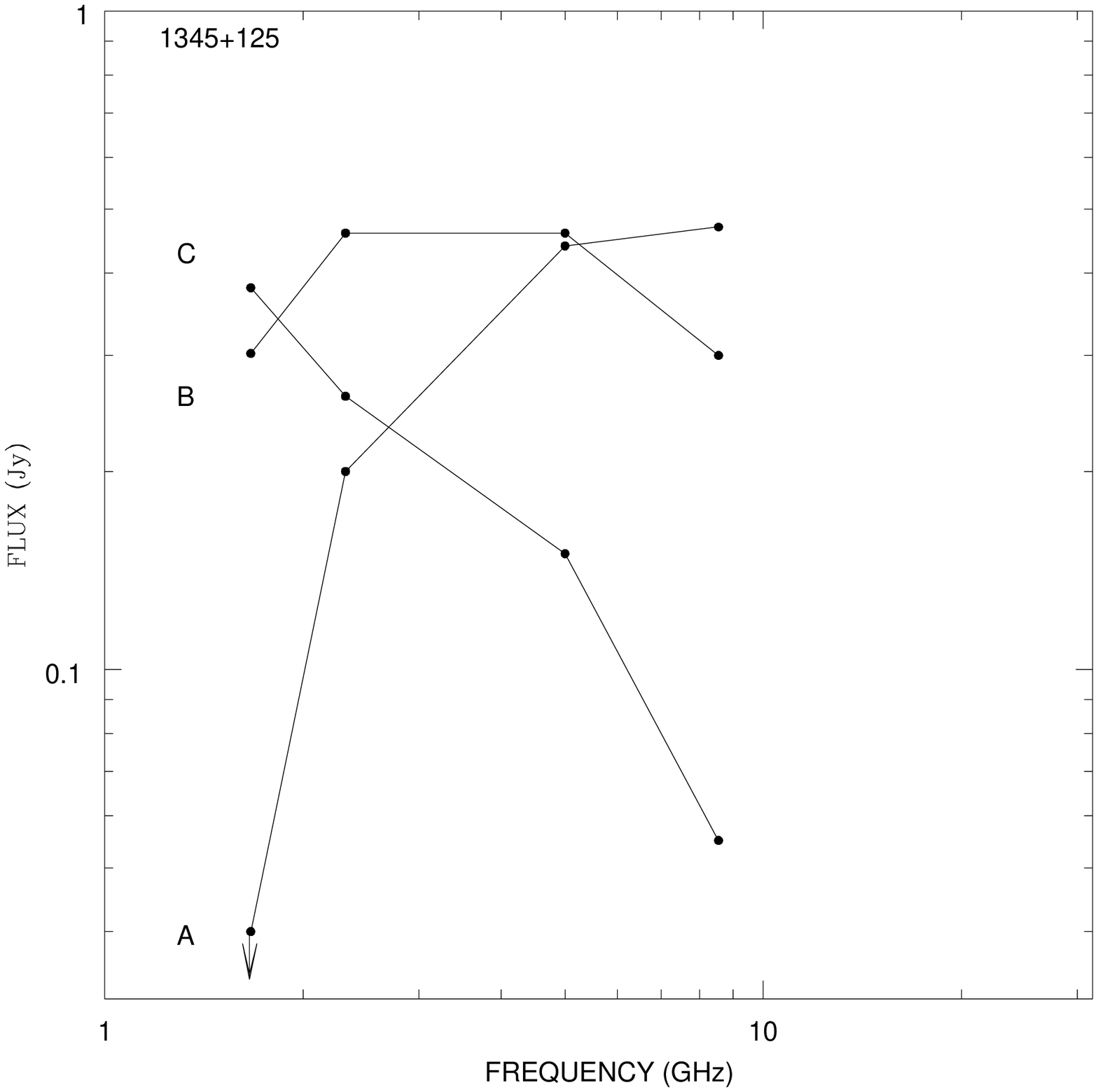}
      \caption{{\bf a)} 1345+125 at 5, 2.27, 1.66 GHz; grey scale image
is full resolution 5 GHz image: the restoring beam is 10$\times$5 mas in p.a. 0$^\circ$, the
r.m.s. noise on the image is 2.5 mJy/beam, the contour levels here and in the following
figures are -3, 3, 6, 12, 25, 50, 100, 200, 400, 800, 1500, 3000 times the r.m.s. noise,
the peak flux density is 630 mJy/beam. {\bf b)} 1345+125 MERLIN image at 1.66 GHz: the restoring
beam is 256$\times$129 mas
in p.a. 37$^\circ$, the r.m.s. noise on the image is 2 mJy/beam, the peak flux density is 4236 mJy/beam.
{\bf c)}   1345+125: spectra of component A, B and C, data from Stanghellini
et al. (1997a), Fey et al. (1996) and this paper.       }
         \label{1345im}
   \end{figure*}
\clearpage
   \begin{figure*}
   \centering
   \includegraphics[width=8cm]{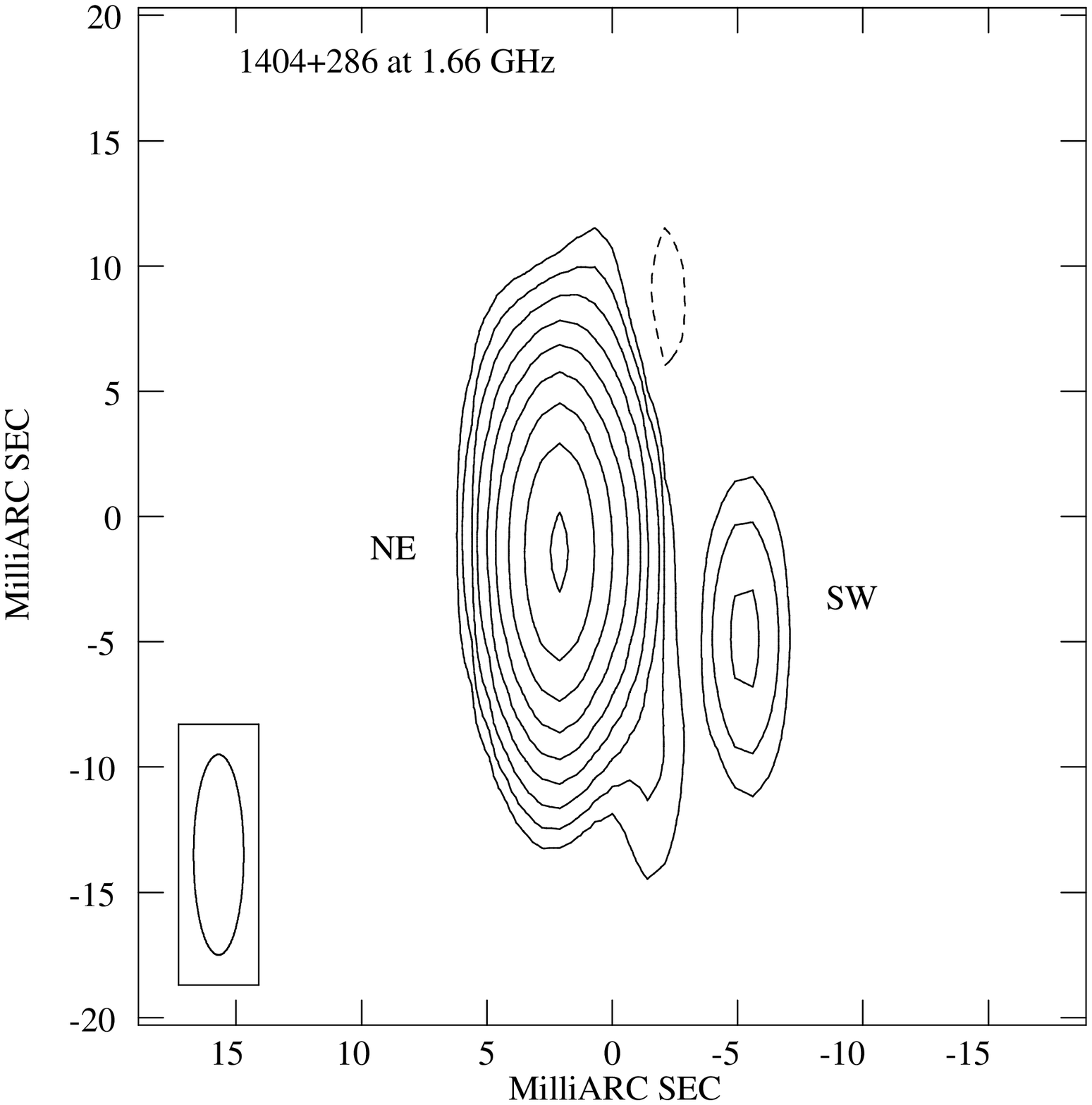}
      \includegraphics[width=8cm]{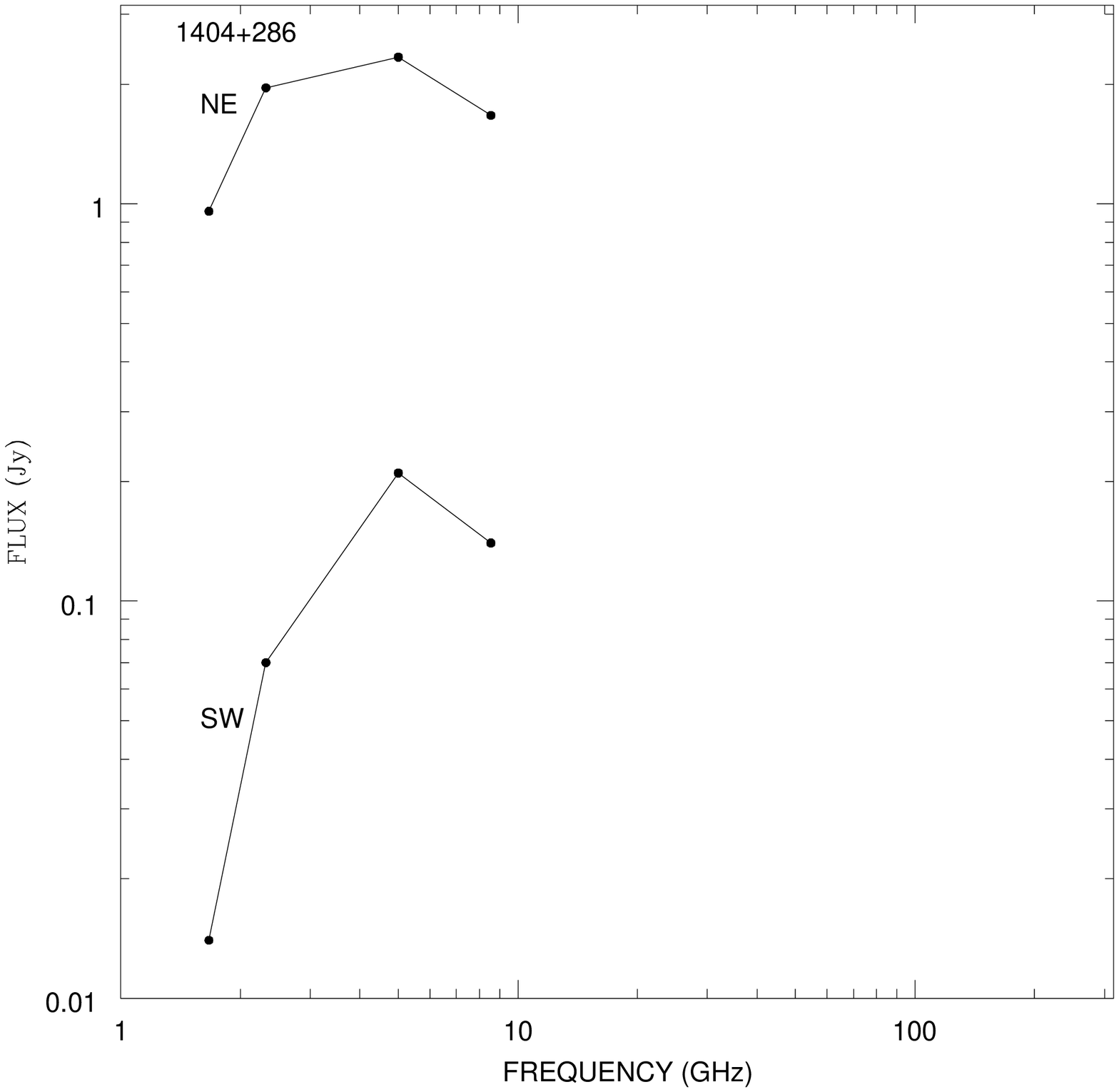}
      \caption{{\bf a)} OQ208 EVN image at 1.66 GHz: the restoring beam is 8$\times$2 mas in p.a. 0$^\circ$, the
r.m.s. noise on the image is 0.5 mJy/beam, the peak flux density is 715 mJy/beam.
 {\bf b)} 1404+286: spectra of component NE and SW, data from Stanghellini
et al. (1997b), Fey et al. (1996) and this paper.            }
         \label{1404im}
   \end{figure*}
   \begin{figure*}
   \centering
   \includegraphics[width=8cm]{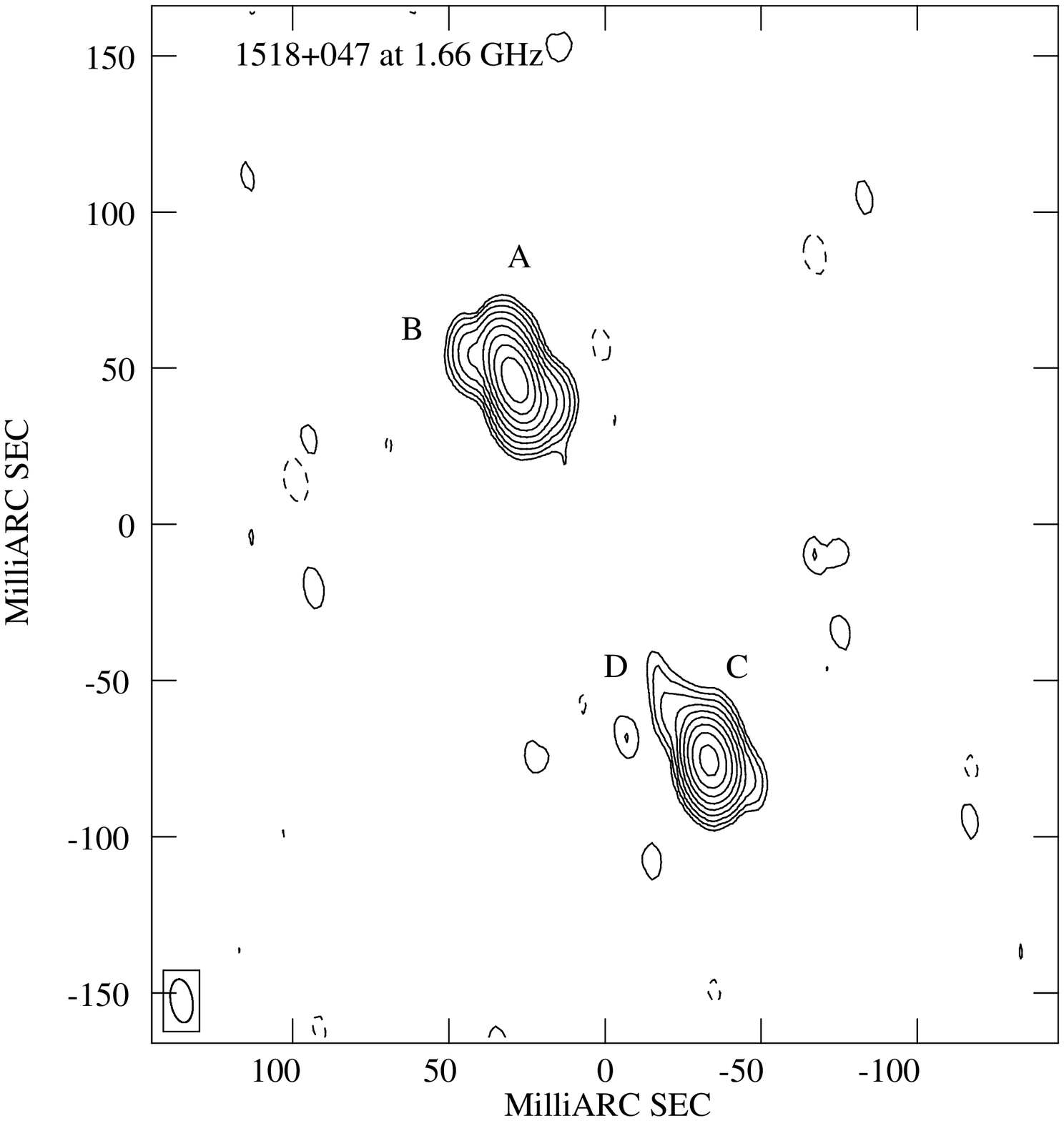}
     \includegraphics[width=8cm]{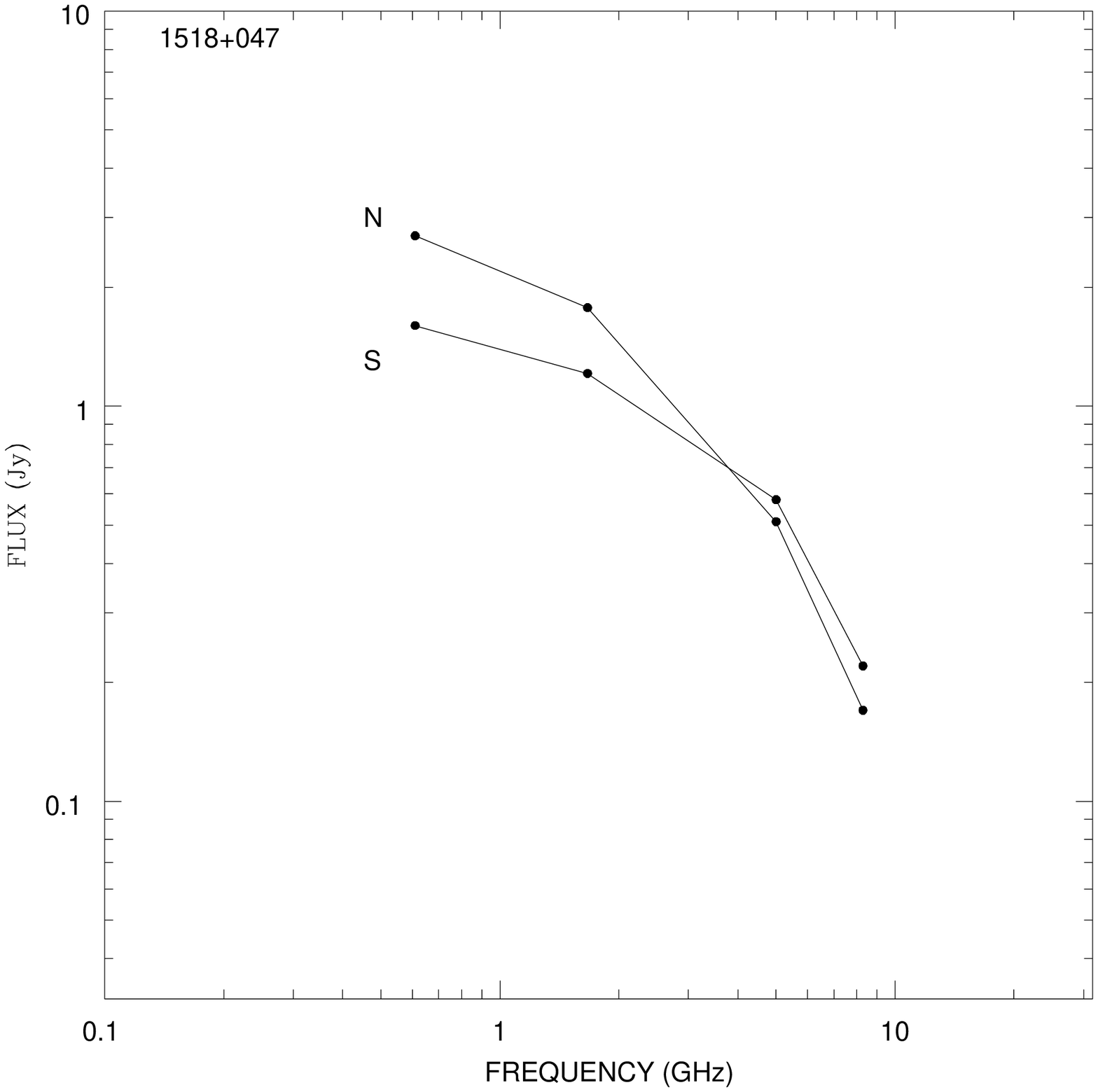}
      \caption{{\bf a)} 1518+047 EVN image at 1.66 GHz: the restoring beam is 14$\times$7 mas in p.a. 8.8$^\circ$, the
r.m.s. noise on the image is 1 mJy/beam, the peak flux density is 1144 mJy/beam.
{\bf b)} 1518+047: spectra of component N and S, data from Mutel et al. (1985),
      Dallacasa et al. (1998), and this paper. Data at 2.3 GHz from Dallacasa et al.
      have not been taken in consideration because a significant
      fraction of the flux density is clearly missing, and were not consistent with the data shown.              }
         \label{1518im}
   \end{figure*}
\clearpage
   \begin{figure}
   \centering
   \includegraphics[width=8cm]{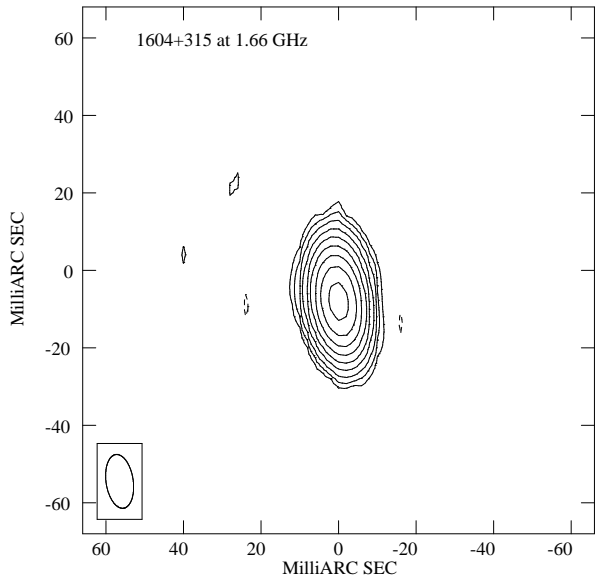}
      \caption{1604+315 EVN image at 1.66 GHz: the restoring beam is 14$\times$7 mas in p.a. 8.4$^\circ$, the
r.m.s. noise on the image is 0.6 mJy/beam, the peak flux density is 662 mJy/beam.
              }
         \label{1604im}
   \end{figure}

   \begin{figure}
   \centering
   \includegraphics[width=8cm]{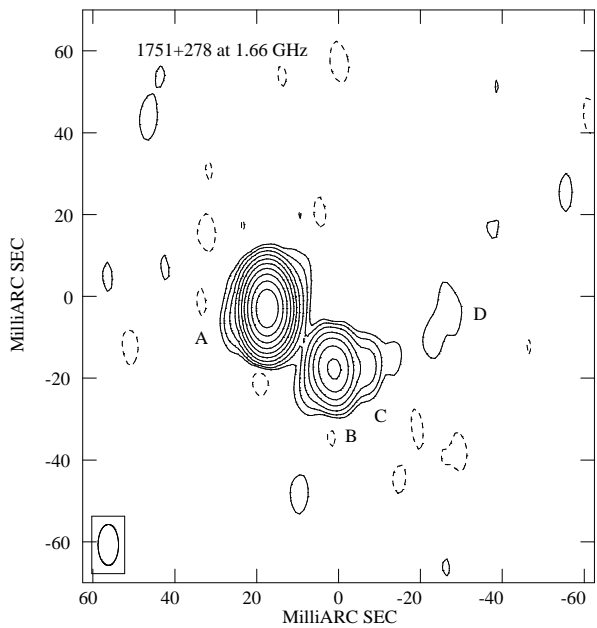}
      \caption{1751+278 EVN image at 1.66 GHz: the restoring beam is 10$\times$5 mas in p.a. 0$^\circ$, the
r.m.s. noise on the image is 0.2 mJy/beam, the peak flux density is 417 mJy/beam.
              }
         \label{1751im}
   \end{figure}

\clearpage
  \begin{figure*}
   \centering
   \includegraphics[width=8cm]{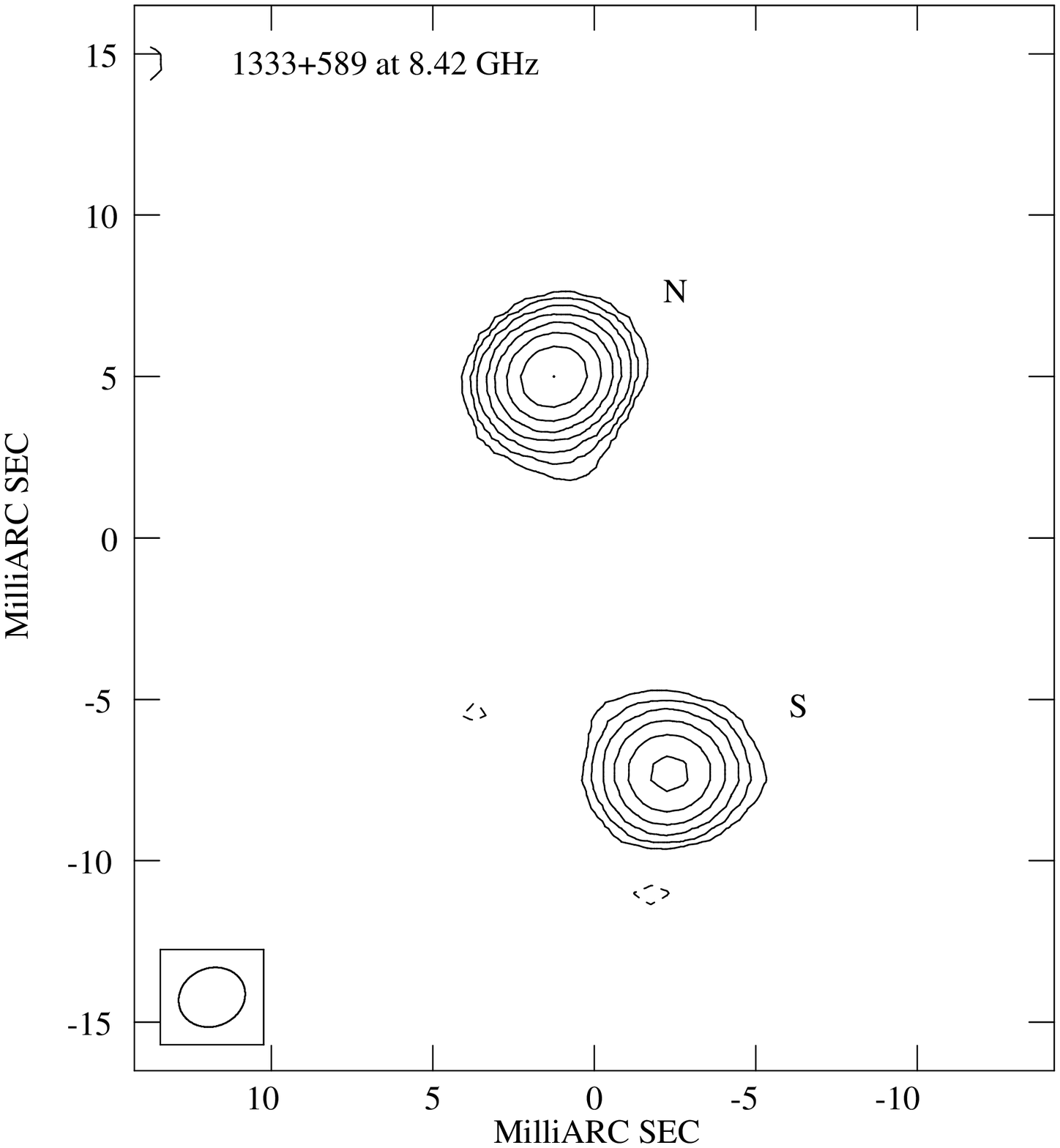}
       \includegraphics[width=8cm]{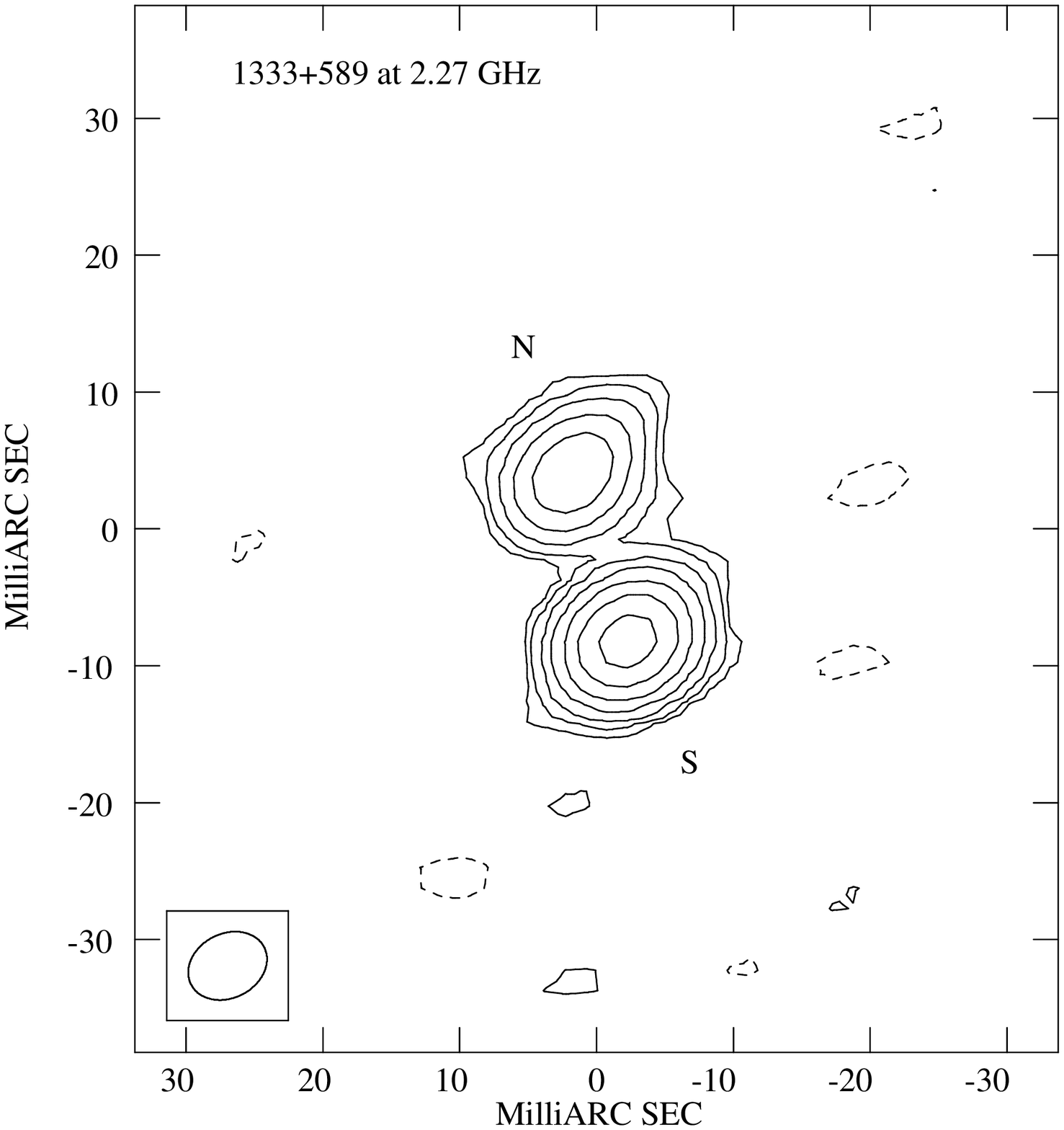}
       \includegraphics[width=8cm]{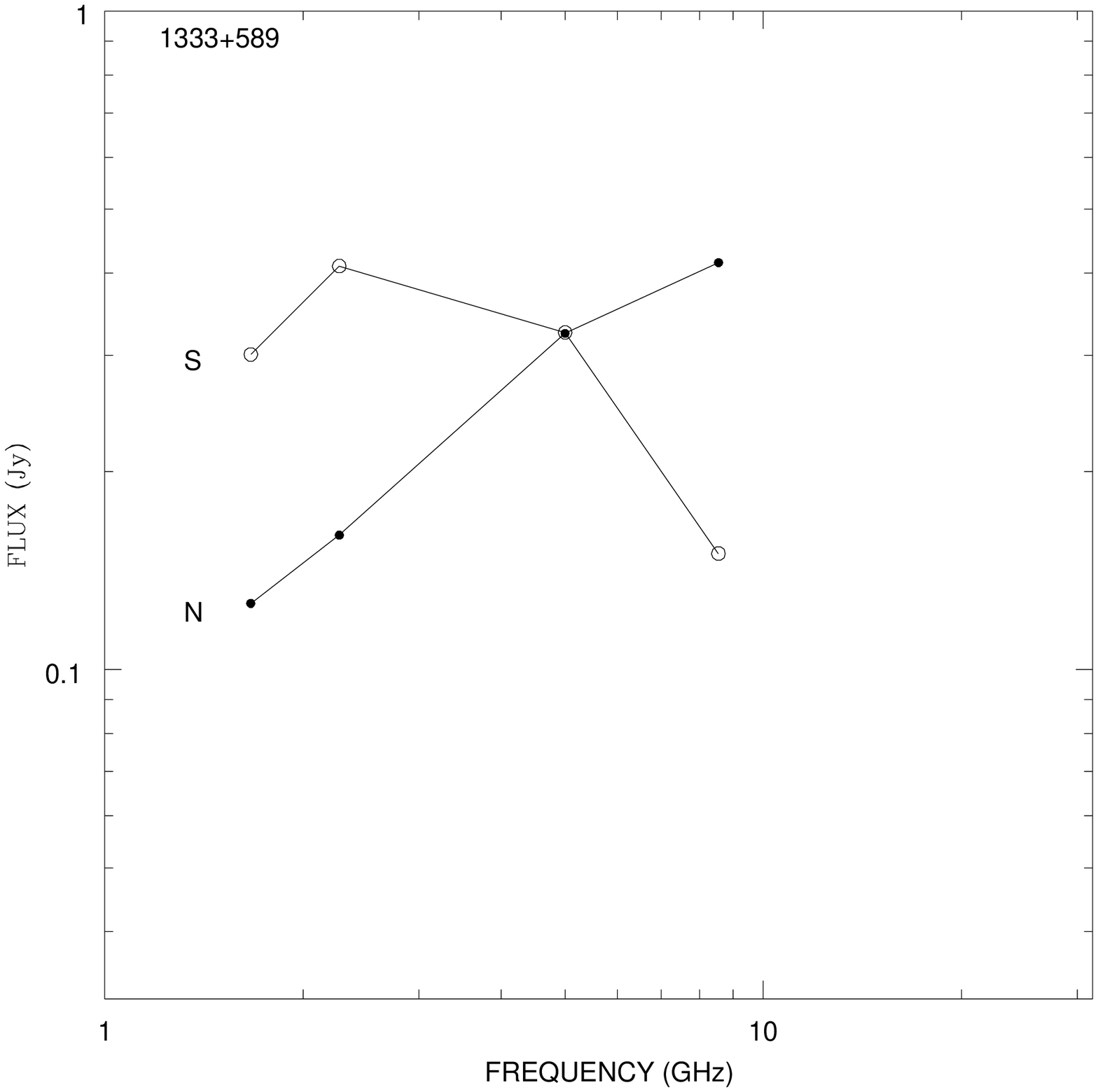}
      \caption{{\bf a)} 1333+589 EVN image at 8.4 GHz: the restoring beam is 2.1$\times$1.8 mas in p.a. 112$^\circ$, the
r.m.s. noise on the image is 1 mJy/beam, the peak flux density is 401 mJy/beam.
{\bf b)} 1333+589 EVN image at 2.3 GHz: the restoring beam is 6$\times$4.7 mas in p.a. 117$^\circ$, the
r.m.s. noise on the image is 1.3 mJy/beam, the peak flux density is 480 mJy/beam.
{\bf c)} 1333+589: spectra of component N, S and N+S, data from Xu et al. (1995), Thakkar et al. (1995)
and this paper.
             }
         \label{1335im}
   \end{figure*}
\clearpage
   \begin{figure}
   \centering
   \includegraphics[width=8cm]{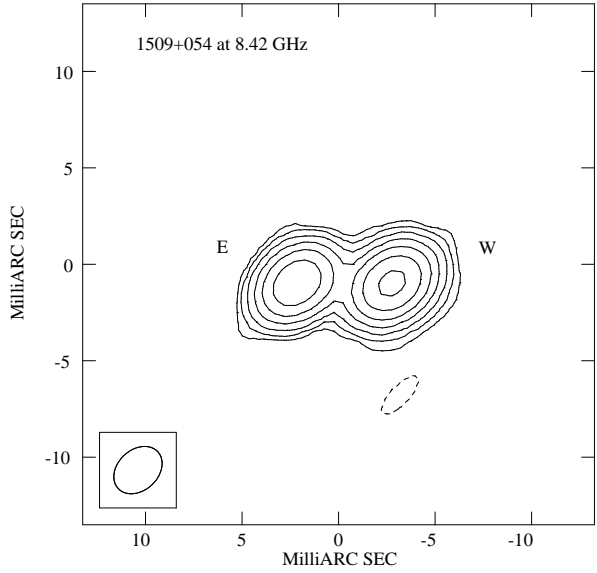}
      \caption{1509+054 EVN image at 8.4 GHz: the restoring beam is 2.8$\times$2.1 mas in p.a. 133$^\circ$, the
r.m.s. noise on the image is 1.5 mJy/beam, the peak flux density is 374 mJy/beam.
              }
         \label{1511im}
   \end{figure}
\begin{figure}
   \centering
   \includegraphics[width=8cm]{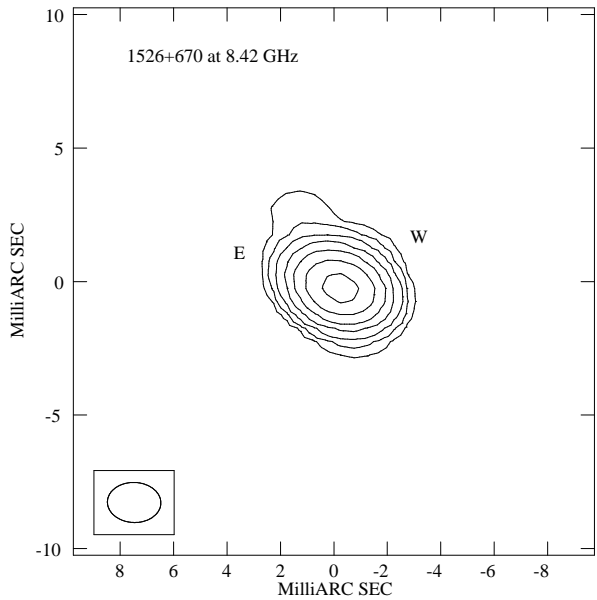}
      \caption{1526+670 EVN image at 8.4 GHz: the restoring beam is 2$\times$1.5 mas in p.a. 88$^\circ$, the
r.m.s. noise on the image is 0.8 mJy/beam, the peak flux density is 216 mJy/beam.
              }
         \label{1526im}
   \end{figure}
      \begin{figure}
   \centering
   \includegraphics[width=8cm]{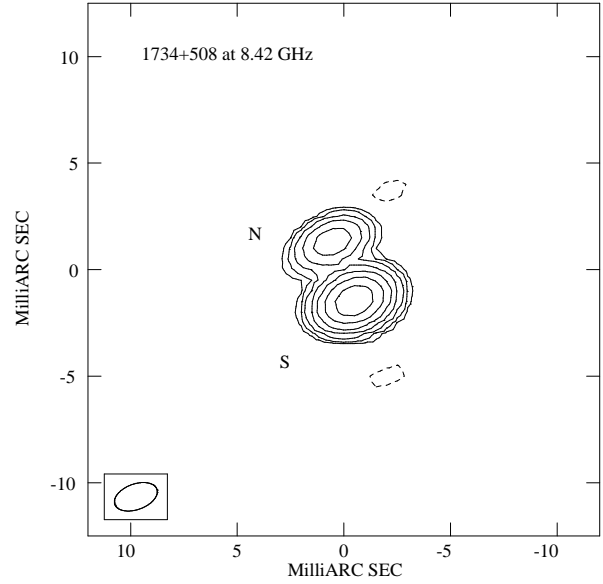}
      \caption{1734+508 EVN image at 8.4 GHz: the restoring beam is 2.1$\times$1.2 mas in p.a. 110$^\circ$, the
r.m.s. noise on the image is 1.5 mJy/beam, the peak flux density is 528 mJy/beam.
              }
         \label{1735im}
   \end{figure}

\end{document}